\documentclass{PoS}

\usepackage{amsmath}
\usepackage{amssymb}

\newcommand{\bc}{\begin{center}}
\newcommand{\ec}{\end{center}}
\newcommand{\be}{\begin{equation}}
\newcommand{\ee}{\end{equation}}
\newcommand{\bea}{\begin{eqnarray}}
\newcommand{\eea}{\end{eqnarray}}
\newcommand{\bi}{\begin{itemize}}
\newcommand{\ei}{\end{itemize}}
\newcommand{\bt}{\begin{tabular}}
\newcommand{\et}{\end{tabular}}

\def\Nf{N_{\rm f}}
 
\def\ks{\kappa_\mathrm{sea}}

\def\mps{m_{\rm PS}}

\def\mN{m_{\rm N}}
\def\ZV{Z_{\rm V}}
\def\cV{c_{\rm V}}
\def\bV{b_{\rm V}}
\def\fm{\;{\rm fm}}
\def\MeV{\;{\rm MeV}}

\def\ks{\kappa_{\rm sea}}
\def\kv{\kappa_{\rm val}}

\title{
\begin{picture}(0,0)(0,0)%
   \put(0,75){\makebox(0,0)[l]{\textnormal{\normalsize DESY 06-183}}}%
   \put(0,60){\makebox(0,0)[l]{\textnormal{\normalsize Edinburgh 2006/26}}}%
\end{picture}%
Nucleon Form Factors: Probing the Chiral Limit}

\ShortTitle{Nucleon Form Factors: Probing the Chiral Limit}

\author{QCDSF/UKQCD Collaboration:
  Meinulf G\"{o}ckeler,$^a$
  Philipp H\"{a}gler,$^b$
  Roger Horsley,$^c$
  Yoshifumi Nakamura,$^d$
  \speaker{Dirk Pleiter},$^d$
  Paul~E.L.~Rakow,$^e$
  Andreas Sch\"{a}fer,$^a$
  Gerrit Schierholz,$^{d,f}$
  Wolfram Schroers,$^d$
  Thomas Streuer,$^g$
  Hinnerk St\"uben$^h$ and
  James M.~Zanotti$^c$\\
  \llap{$^a$} Institut f\"ur Theoretische Physik, Universit\"at Regensburg,
              93040 Regensburg, Germany\\
  \llap{$^b$} Institut f\"ur Theoretische Physik T39, Physik-Department der
              TU M\"unchen, 85747~Garching, Germany\\
  \llap{$^c$} School of Physics, University of Edinburgh, Edinburgh EH9 3JZ,
              UK\\
  \llap{$^d$} John von Neumann-Institut f\"ur Computing NIC, Deutsches
              Elektronen-Synchrotron DESY, 15738 Zeuthen, Germany\\
  \llap{$^e$} Theoretical Physics Division, Department of Mathematical Sciences,
              University of Liverpool, Liverpool L69 3BX, UK\\
  \llap{$^f$} Deutsches Elektronen-Synchrotron DESY, 22603 Hamburg, Germany\\
  \llap{$^g$} Department of Physics and Astronomy, University of Kentucky,
              Lexington KY 40506, USA\\
  \llap{$^h$} Konrad-Zuse-Zentrum f\"ur Informationstechnik Berlin,
              14195 Berlin, Germany\\
  Email: \email{dirk.pleiter@desy.de}
}

\abstract{
The electromagnetic form factors provide important hints for the internal
structure of the nucleon and continue to be of major interest for
experimentalists. For an intermediate range of momentum transfers the form factors
can be calculated on the lattice.  However, reliability of the results
is limited by systematic errors due to the required extrapolation to
physical quark masses. Chiral effective field theories predict a rather
strong quark mass dependence in a range which was yet unaccessible for
lattice simulations. We give an update on recent results from the QCDSF
collaboration using gauge configurations with $\Nf=2$, non-perturbatively
O(a)-improved Wilson fermions at very small quark masses down to 340
MeV pion mass, where we start to probe the relevant quark mass region.
}

\FullConference{XXIV International Symposium on Lattice Field Theory\\
		July 23-28 2006\\
		Tucson Arizona, US}

\begin{document}

\section{Introduction}

In recent years the phenomenological interest in the electromagnetic
form factors of the nucleon has revived. This was triggered by the
Jefferson Lab polarisation experiments \cite{JLabHallA,Punjabi:2005wq}
measuring the ratio of the proton electric to magnetic form factors,
$\mu^{\rm (p)}G_e^{\rm (p)}(Q^2)/G_m^{\rm (p)}(Q^2)$.
From these measurements an unexpected decrease of this ratio has
been found, which means that the proton's electric form factor falls
off faster than the magnetic form factor.

Many theoretical calculations have been done to investigate possible
interpretations of a decrease of this ratio (see, e.g.,
\cite{Punjabi:2005wq} for an overview). Lattice techniques
allow the calculation of the form factors from first principles.
Such calculations do not only yield
phenomenologically interesting quantities such as magnetic and electric
charge radii and magnetic moments. These techniques also allow, e.g.,
the investigation of the $Q^2$ dependence of the nucleon electromagnetic
form factors, which can be compared with experimental results and also
helps in understanding the asymptotic behaviour of these form factors.

In practice, the calculation of these form factors on the lattice remains
a challenge.  In recent years progress has been made to improve control
on systematic errors which are related to the fact that the calculations
are performed on finite volumes, at finite lattice spacings and at quark masses
which are still relatively large.  For making reliable predictions at physical
quark masses it turned out that numerical results at smaller quark masses are
crucial.  With recent advances in computing power available to lattice QCD
calculations and the speed-up of algorithms for simulating dynamical fermions
it is now possible to reach much smaller quark masses with pseudoscalar
masses in the range of 300 MeV.

\section{Calculation details}

In this talk we present results obtained on configurations with two
mass degenerate flavours of non-perturbatively $O(a)$-improved Wilson
fermions. We choose Wilson glue for the gauge action.
To scale the lattice results for different $\beta$ and $\ks$
we use the Sommer parameter $r_0(\beta,\ks)/a$
and the conversion factor
$r_0 = 0.467 \fm$ to translate our results into physical units.

The form factors are obtained from the standard decomposition of the
nucleon electromagnetic matrix elements
\be
\langle p',s' | J^\mu | p,s \rangle =
\overline{u}(p',s')
\left[
  \gamma_\mu F_1(Q^2) + i\sigma^{\mu\nu}\frac{q_\nu}{2 M_N} F_2(Q^2)
\right]
u(p,s),
\ee
where $p$ ($s$) and $p'$ ($s'$) denote initial and final momenta
(spins), $q = p' - p$ the momentum transfer (with $Q^2=-q^2$) and
$M_N$ the nucleon mass. By calculating the matrix elements on the
l.h.s. and the nucleon mass we obtain the Dirac
form factor $F_1(Q^2)$ and Pauli form factor $F_2(Q^2)$.

The nucleon matrix elements are extracted from ratios of three- and
two-point functions:
\be
 R(t,\tau,\vec{p}\,',\vec{p}\,) =
\frac{C_3(t,\tau,\vec{p}\,',\vec{p}\,)}{C_2(t,\vec{p}\,')} \times
 \left[ \frac{C_2(\tau,\vec{p}\,') C_2(t,\vec{p}\,') C_2(t-\tau,\vec{p}\,) }
{C_2(\tau,\vec{p}\,) C_2(t,\vec{p}\,) C_2(t-\tau,\vec{p}\,')}
\right]^{1/2}.
\label{eq:ratio}
\ee
%
%
Here $t$ denotes the location of the sink. Assuming the source being
located at time slice $0$, we expect an plateau for
$0 \ll \tau \ll t$ (source and sink are separated by
a distance of $\sim 1.1$ fm). For further details see
\cite{Gockeler:2003ay}.
We use three different polarisations
$\Gamma_{\rm unpol} = \frac{1}{2}(1+\gamma_4)$,
$\Gamma_1 = \frac{1}{2}(1+\gamma_4)\, i\gamma_5\gamma_1$,
$\Gamma_2 = \frac{1}{2}(1+\gamma_4)\, i\gamma_5\gamma_2$
as well as three different sink momenta
$\vec{p}_0 = ( 0, 0, 0 )$,
$\vec{p}_1 = ( p, 0, 0 )$,
$\vec{p}_2  = ( 0, p, 0 )$
(where $p=2\pi/L_S$). 17 different choices for the momentum transfer
$\vec{q}$ have been used.
Due to statistical fluctuations the operand of the square root
in Eq.~(\ref{eq:ratio}) may become negative. Results for which this happens
are discarded from the consecutive analysis steps.

We use the local vector current, which needs to be renormalised
and improved:
\vspace*{-1mm}
\be
 V_\mu = \ZV (1+\bV a m_q) \left[ \bar{q} \gamma_\mu q
   + {\mathrm i} \cV a \partial_\lambda
      ( \bar{q} \sigma_{\mu \lambda} q ) \right].
\ee
The renormalisation coefficient $\ZV$ and the parameter $\bV$ have been
determined non-perturbatively \cite{qcdsf-Z}. The improvement coefficient
$\cV$ is only known perturbatively. However, since this coefficient is
expected to be a small number and because the improvement term was
found to be small in the quenched approximation
\cite{Capitani:1998ff}, we will ignore the improvement of the operator.

In the following we will consider the iso-vector, iso-scalar and the proton
form factors. The latter two might receive contributions from quark-line
disconnected
terms, which are notoriously hard to calculate on the lattice and will not
be considered here.

\begin{figure}[t]
\includegraphics[scale=0.4,angle=-90]{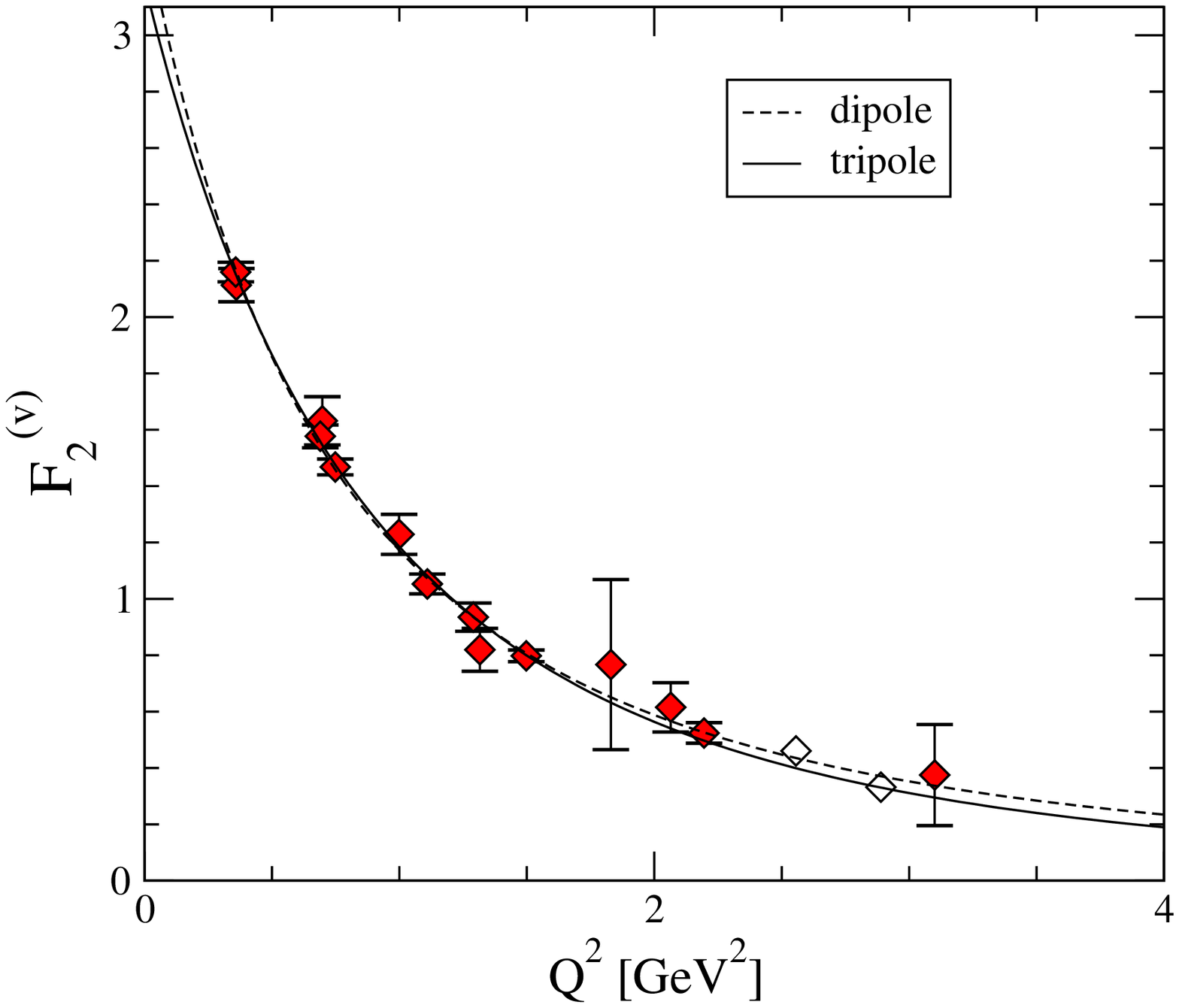} \hfill
\includegraphics[scale=0.4,angle=-90]{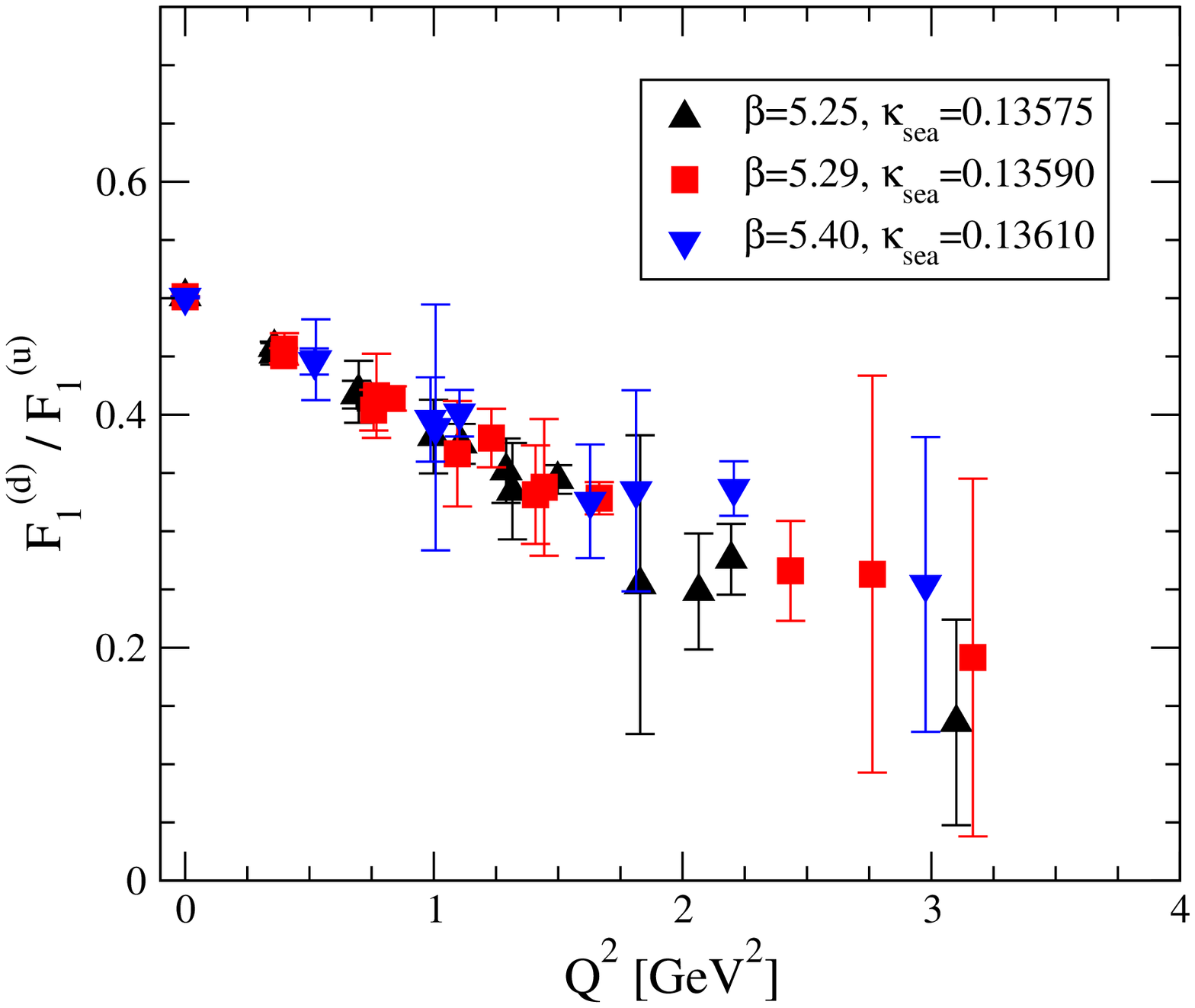}
\caption{\label{fig:q2scaling}
The iso-vector Pauli form factor radius at $(\beta,\ks)=(5.25, 0.13575)$
(left plot) and the ratio of the iso-vector $d$- and $u$-flavour Dirac form factors
(right) as a function of $Q^2$.
In the right plot we show results for similar $\mps \simeq 400 \MeV$ but
different lattice spacings.}
\vspace*{-2mm}
\end{figure}

\section{Parameterisation and $Q^2$ dependence}

First we will investigate the $Q^2$ dependence of the Dirac
and Pauli form factors. Both lattice and experimental data can be reasonably
well parametrised by a pole ansatz
\vspace*{-1mm}
\be
F_i(Q^2) = \frac{F_i(0)}{(1+Q^2/M_i^2)^p}.
\label{eq:poleansatz}
\vspace*{-1mm}
\ee

From naive dimensional counting one would expect the Dirac and Pauli form
factors to scale differently, i.e. $F_1 \propto Q^{-4}$ and $F_2 \propto Q^{-6}$,
which corresponds to $p=2$ and $p=3$, respectively. Experimental data as well as
theoretical calculations indicate deviations from this naive picture.
For instance, the JLab results for $\sqrt{Q^2} F_2^{\rm (p)}(Q^2)/F_1^{\rm (p)}(Q^2)$
were found to be surprisingly flat. A perturbative QCD analysis of
the Pauli and Dirac form factors predicts the ratio
\be
(Q^2/log^{2+8/(9\beta)} Q^2/\Lambda^2) F_2(Q^2)/F_1(Q^2)
\ee
(with $\beta=11 - 2\Nf/3$) to scale as a constant \cite{Belitsky:2002kj}.
In an investigation of the $Q^2$ dependence of the experimental nucleon
form factor data using empirical parameterisations, Diehl and collaborators 
found indications for the form factor scaling to be flavour dependent
\cite{Diehl:2004cx}. By comparing the Dirac form factor
results for proton and nucleon they found the flavour contribution
$F_1^{\rm (d)}(Q^2)$ to decrease faster with $Q^2$ than $F_1^{\rm (u)}(Q^2)$.

From a first inspection of the lattice results one finds that the effects
from changing $p$ in Eq.~(\ref{eq:poleansatz}) are small with respect to the
statistical errors. In Fig.~\ref{fig:q2scaling} (left plot) we show the results
of a fit to $F_2^{\rm (v)}$ for one particular data set using $p=2$ and $3$.
It is clear that it is difficult to obtain lattice data with high enough
precision over a large enough range of $Q^2$ values to distinguish between
a dipole or tripole behaviour. It may, however, be instructive to consider
ratios of form factors in order to reveal significant deviations from the naive
scaling hypothesis. In the right plot of Fig.~\ref{fig:q2scaling} we show the
ratio $F_1^{\rm (d)}/F_1^{\rm (u)}$ and find that it does not scale as a constant.
This is consistent
with the observation by Diehl et al. We should however emphasize
that we are ignoring possible contributions from disconnected terms.

Based on the above observation we perform fits to the Dirac and Pauli form factors
for each flavour separately using the ansatz Eq.~(\ref{eq:poleansatz}).
To eventually decide which $p$ should be used, we perform these fits for
various $1.5 \lesssim p \lesssim 4$ and search for the ``optimal'' $p$
for which $\chi^2/\mathrm{d.o.f.}$ becomes minimal.
We observe strong variations of the resulting
``optimal'' $p$ for data sets which differ in ($\beta$,$\ks$,$\kv$).
However, we nevertheless see clear trends.
In case of the $u$-flavour Dirac form factor $F_1^{\rm (u)}(Q^2)$ we find for most
data sets $p$ to be close to 2. On the other hand, for the other form factors
$F_1^{\rm (d)}(Q^2)$, $F_2^{\rm (u)}(Q^2)$ and $F_2^{\rm (d)}(Q^2)$ we typically get
$p \simeq 3$.

\section{Form factor radii and magnetic moments}

From the fits to Eq.~(\ref{eq:poleansatz}) we obtain the form factors at
zero momentum transfer $F_i(0)$ and the di- or tripole masses $M_i$.
Equivalently, we can use the same fit to obtain the form factor radii $r_i$
and the magnetic moment $\mu$. These quantities are defined as follows:
\bea
F_i(Q^2) &=& F_i(0) \left[ 1 - \frac{1}{6} \; {r_i^2} Q^2
                               + {\mathcal O}(Q^4) \right], \\
\mu &=& F_1(0) + F_2(0).
\eea
From the magnetic moment we can calculate the anomalous magnetic moment, e.g.
$\kappa^{\rm (v)} = \mu^{\rm (v)} - 1$. For comparison with phenomenological results
we use the normalised anomalous magnetic moment, e.g.
%
$\kappa^{\rm (v){\rm norm}} = \kappa^{\rm (v)}\;\mN(m_\pi)/\mN(\mps)$,
%
where $\mN(\mps)$ refers to the nucleon mass calculated on the lattice
at the quark mass corresponding to the pseudoscalar mass $\mps$
and $\mN(m_\pi)$ the experimental value of the nucleon mass.

\begin{figure}[t]
\includegraphics[scale=0.4,angle=-90]{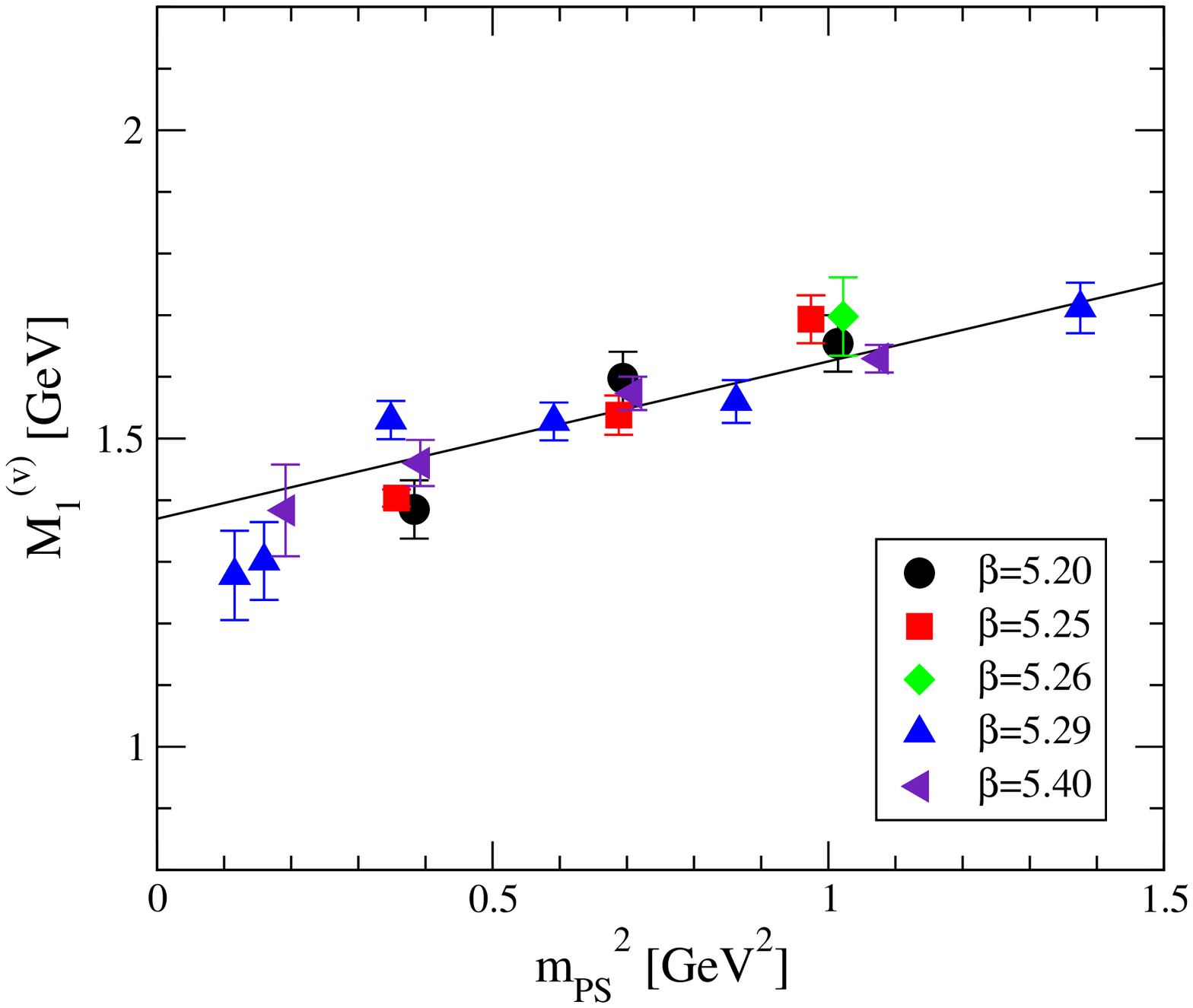} \hfill
\includegraphics[scale=0.4,angle=-90]{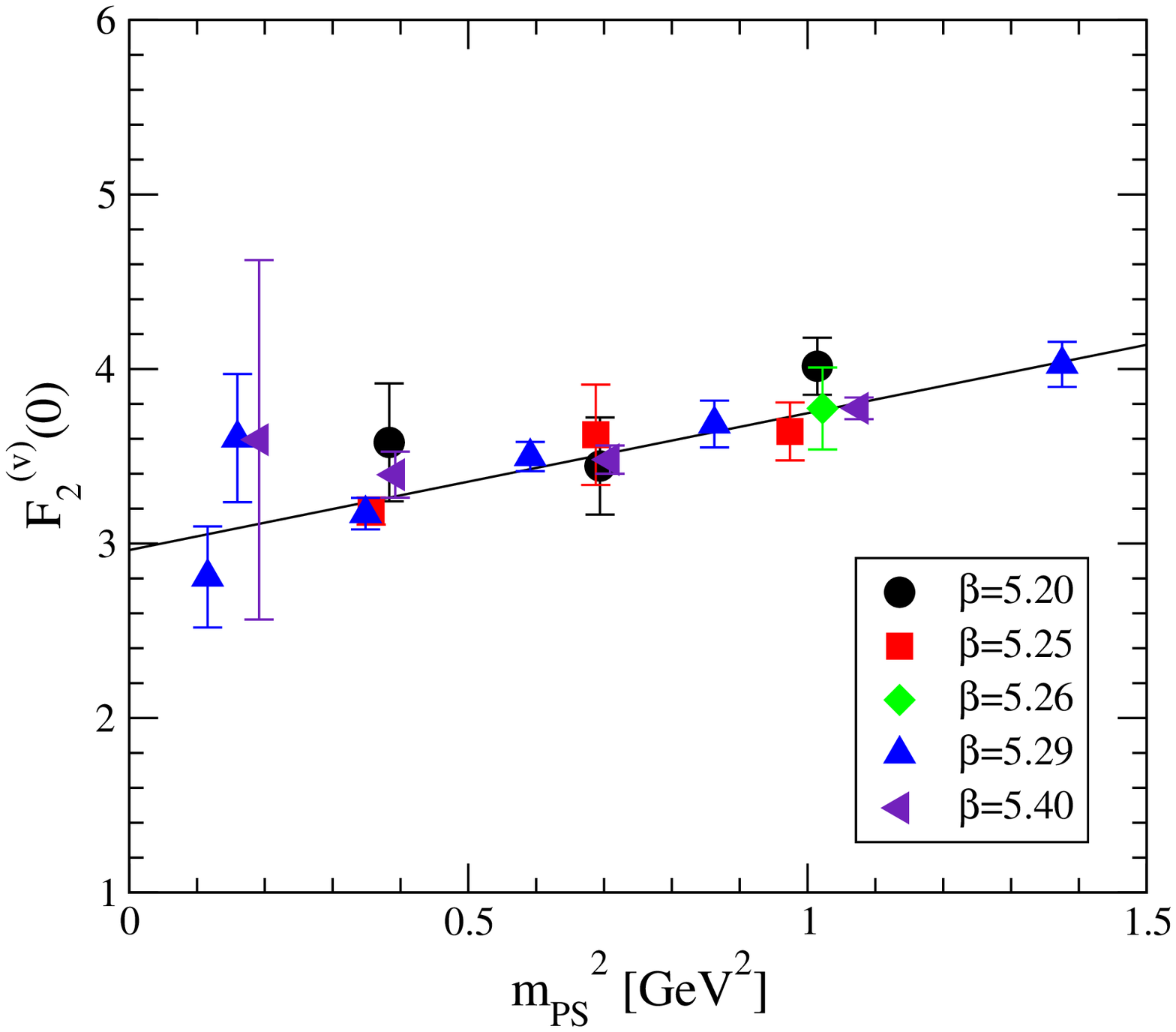}
\caption{\label{fig:f2}
The iso-vector Dirac form factor dipole mass (left plot) and the
iso-vector Pauli form factor at zero momentum transfer (right plot)
as a function of $\mps^2$.}
\vspace*{-2mm}
\end{figure}

In Fig.~\ref{fig:f2} we show results for $M_1^{\rm (v)}$ and $F_2^{\rm (v)}(0)$
as a function of $\mps^2$ which have been calculated
for different values of the gauge coupling $\beta$ and various sea quark 
masses. These results seem to lie on a universal curve, which indicates
that discretisation errors are small. In the following we will consider
them as negligible
compared to the statistical errors. We furthermore observe that the
iso-vector anomalous magnetic moments and di-/tripole masses show a
linear quark mass dependence for a very large range of quark masses.
However, using an ansatz linear in $(r_0 \mps)^2$ to extrapolate our results
to the physical point, we obtain values which differ significantly from
those extracted from experiment. This is consistent with calculations based on
a chiral effective field theory (ChEFT) that includes nucleons, pions and delta
resonances as explicit degrees of freedom
\cite{Hemmert:2002uh,Gockeler:2003ay,Leinweber:2001ui}.
These calculations predict for the
iso-vector form factor radii and the iso-vector anomalous magnetic moment
a strong quark mass dependence in the small quark mass region. This region
is just starting to become accessible for simulations with dynamical Wilson
quarks.

\begin{figure}[t]
\includegraphics[scale=0.4,angle=-90]{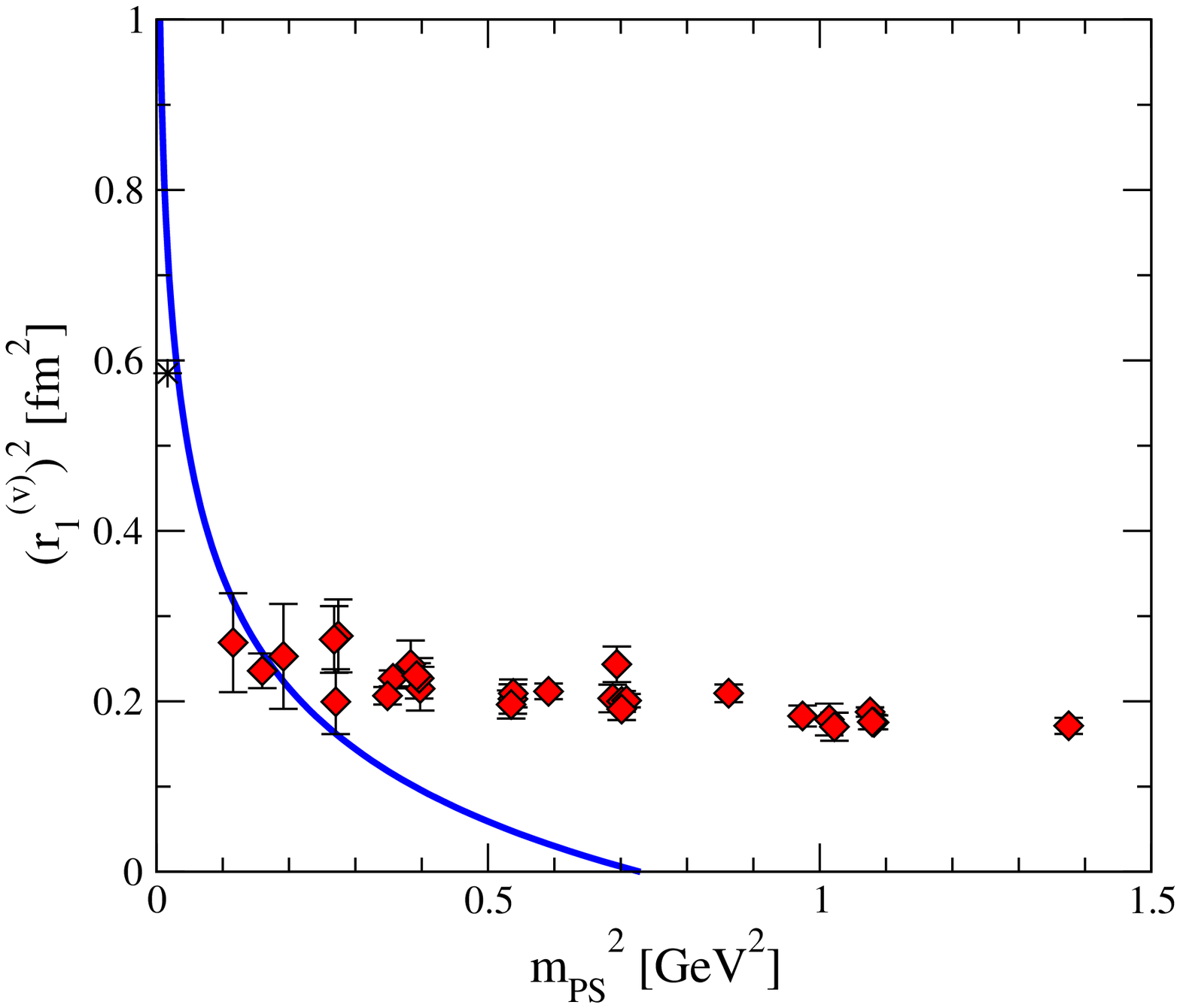} \hfill
\includegraphics[scale=0.4,angle=-90]{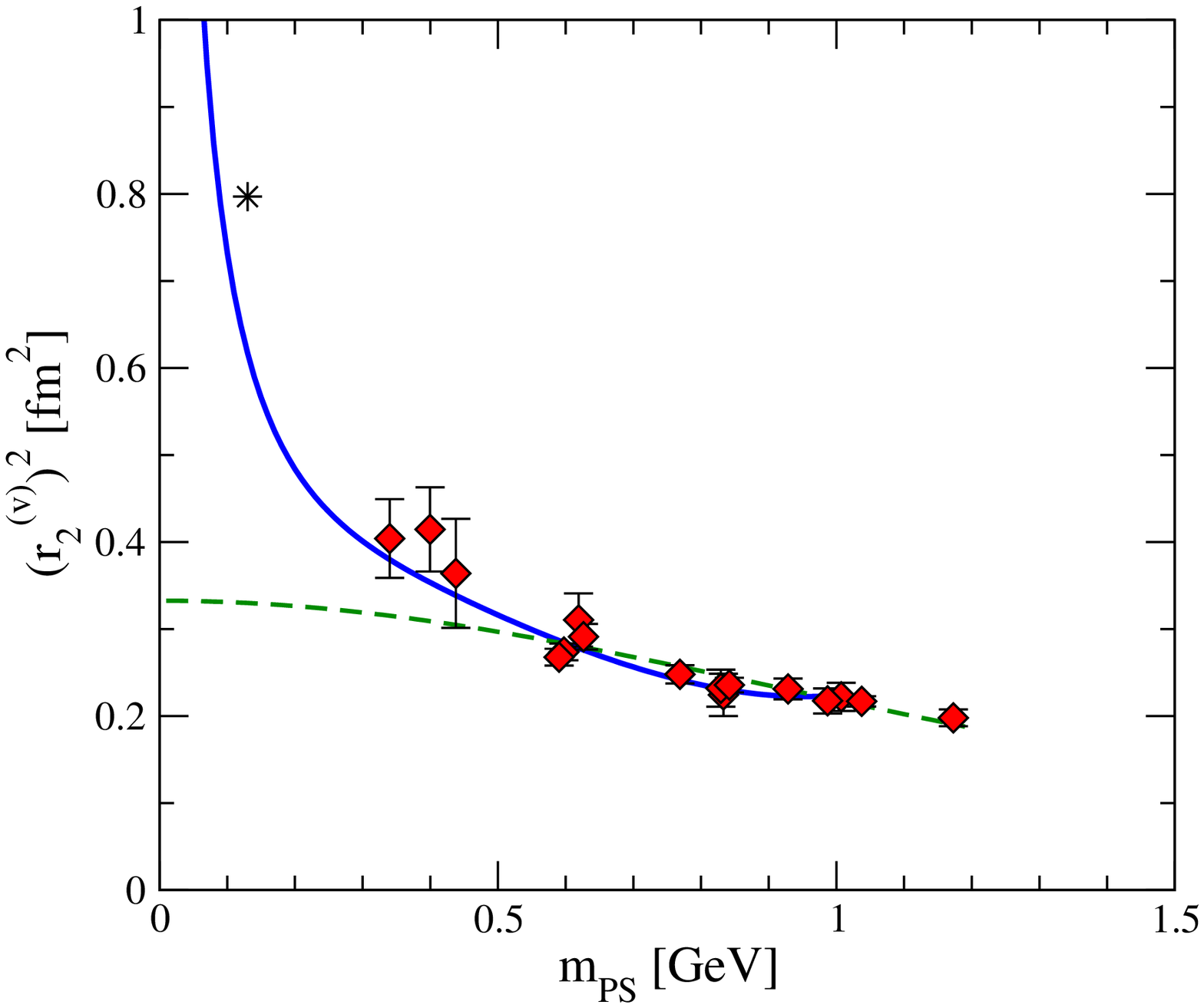}
\caption{\label{fig:radii}
Lattice results for the Dirac (left) and Pauli (right)
radii. The solid lines show the ChEFT results
given in Eqs.~(4.3) and (4.4). Note that the left
curve is not based on a fit. In the right-hand plot the dashed line is
the result of a fit to the tripole masses linear in $(r_0 \mps)^2$.
The star denotes the experimental value.}
\vspace*{-2mm}
\end{figure}

In the left plot of Fig.~\ref{fig:radii} we compare the lattice results for
the iso-vector Dirac form factor radius
with the following result from ChEFT, where we used the same phenomenological
parameters as in \cite{Gockeler:2003ay}:
\bea
\label{eq:r1v}
 \left(r_{1}^{\rm (v)}\right)^2 &=&
    -  \frac{1}{(4\pi F_\pi)^2}\left\{1+7 g_{A}^2 +
  \left(10 g_{A}^2 +2\right)
  \log\left[\frac{\mps}\lambda\right]\right\} \\\nonumber
&&     +  \frac{{c_A}^2}{54\pi^2 F_{\pi}^2}\Bigg\{
        26+30\log\left[\frac{\mps}{\lambda}\right]
          +30\frac{\Delta}{\sqrt{\Delta^2-\mps^2}}
             \log\left[\frac{\Delta}{\mps}
            +\sqrt{\frac{\Delta^2}{\mps^2}-1}\right] \Bigg\}.
\eea

\begin{figure}[b]
\includegraphics[scale=0.4,angle=-90]{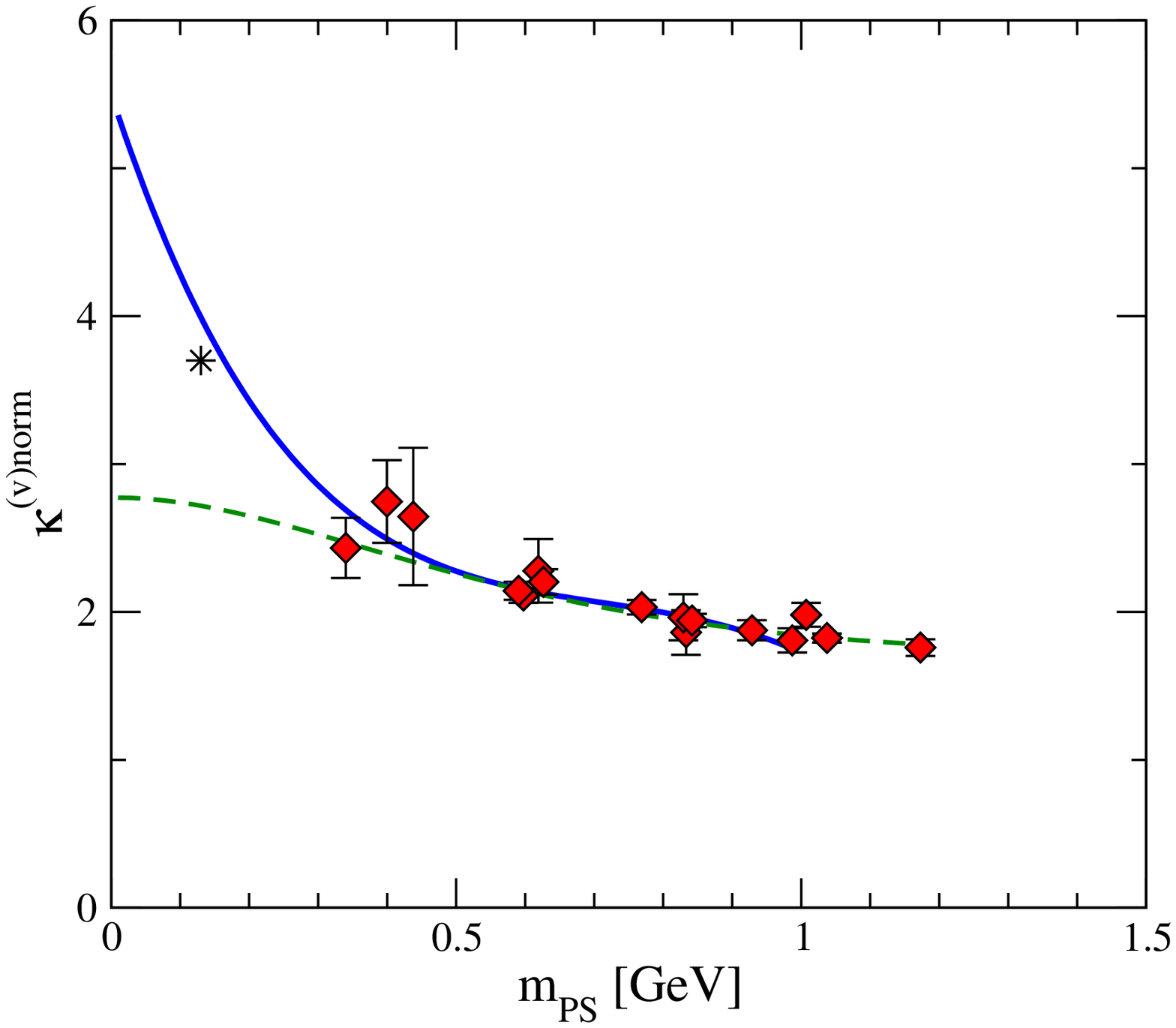} \hfill
\includegraphics[scale=0.4,angle=-90]{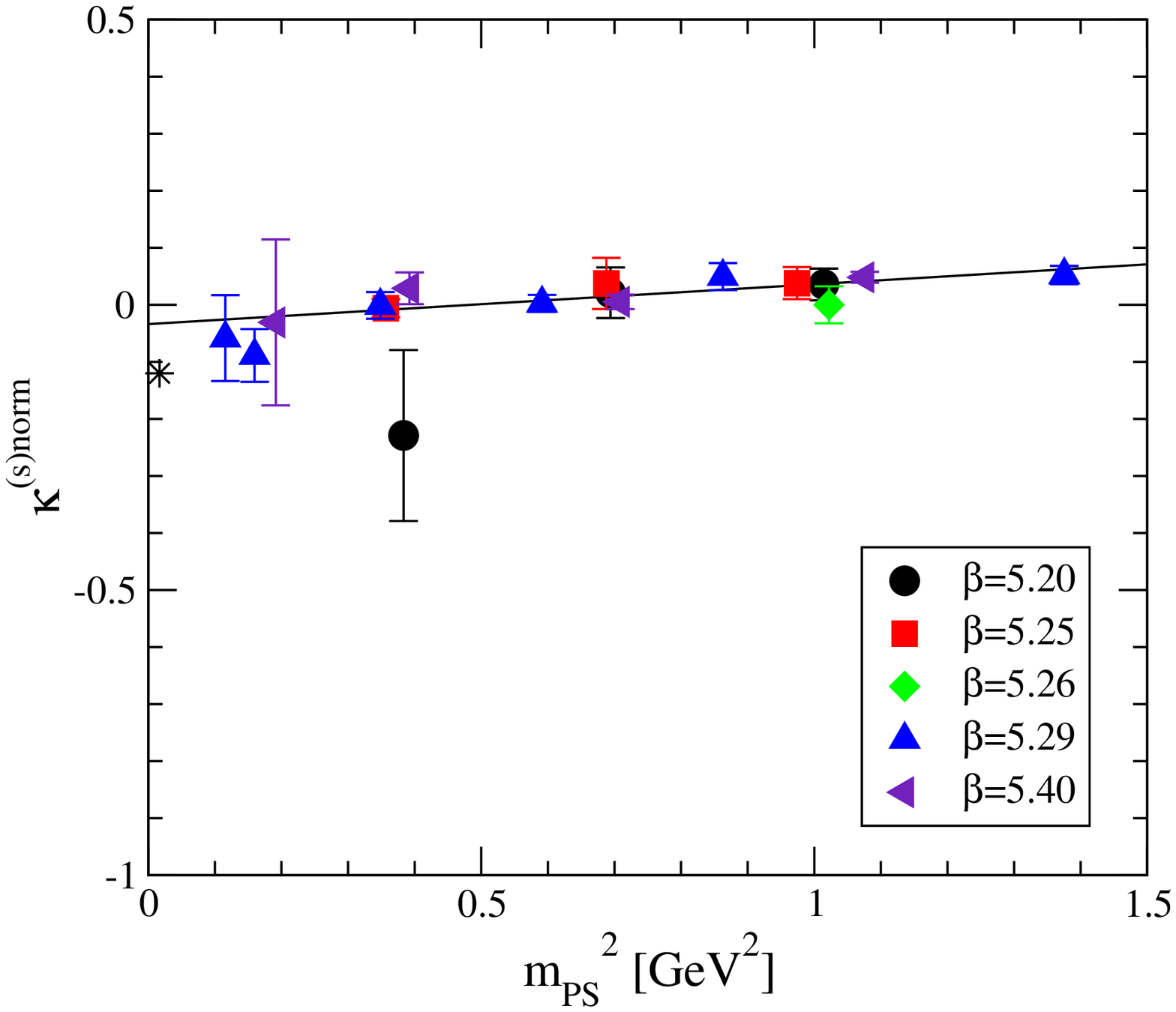}
\caption{\label{fig:mmoment}
Lattice results for the iso-vector (left) and iso-scalar (right)
anomalous magnetic moment. The solid lines in the left plot show the
result from a fit to the ChEFT expression in Eq.~(4.5).
In the right-hand plot an ansatz linear in $(r_0 \mps)^2$ has been
used to fit the data.
The star denotes the phenomenological value.}
\end{figure}

For the iso-vector Pauli form factor, $r_{2}^{\rm (v)}$,
and the anomalous magnetic moment, $\kappa^{\rm (v)}$, we
performed a combined fit of the lattice results to the following expressions
from ChEFT:
\bea
\label{eq:r2v}
\left(r_{2}^{\rm (v)}\right)^2 &=&
   \frac{{g_{A}}^2 {M_N}}
        {8 {F_{\pi}}^2 \kappa^{\rm (v)} (\mps) \pi \mps} + \\\nonumber
&& \frac{{c_A}^2 M_N}{9 F_{\pi}^2 \kappa^{\rm (v)} (\mps)
                  \pi^2 \sqrt{{\Delta}^2-m_{\pi}^2}}
    \log\left[\frac{\Delta}{\mps}+\sqrt{\frac{\Delta^2}{m_{\pi}^2}-1}\right]
  + \frac{24 M_N}{\kappa^{\rm (v)}(\mps)}\,{B_{c2}},
\eea
%
\bea
\label{eq:kv}
 \kappa^{\rm (v)} (\mps) &=&
{\kappa^{{\rm (v)}0}}-\frac{{g_A}^2\,\mps {M_N}}{4\pi {F_\pi}^2} +
\frac{2 {c_A}^2 {\Delta} M_N}{9\pi^2 F_\pi^2}
  \left\{\sqrt{1-\frac{\mps^2}{\Delta^2}}\log R(\mps) +
  \log\left[\frac{\mps}{2\Delta}\right] \right\}
\\\nonumber &&
        -   8 {E_1^{(r)} (\lambda)} M_N \mps^2
        +  \frac{4c_A {c_V} g_A M_N \mps^2}{9\pi^2 F_\pi^2}
         \log\left[\frac{2\Delta}{\lambda} \right]
        +  \frac{4c_A c_V g_A M_N \mps^3}{27\pi F_\pi^2\Delta}
\\\nonumber &&
         -   \frac{8 c_A c_V g_A \Delta^2 M_N}{27\pi^2 F_\pi^2}
         \Bigg\{\left(1-\frac{\mps^2}{\Delta^2}\right)^{3/2} \log R(\mps)
          + \left(1-\frac{3\mps^2}{2\Delta^2}\right)
           \log\left[\frac{\mps}{2\Delta}\right] \Bigg\}.
\eea
Here we keep the chiral limit of the anomalous magnetic moment $\kappa^{{\rm (v)}0}$,
the iso-vector $N$-$\Delta$ coupling $c_V$ and the ChEFT parameters
$B_{c2}$ and $E_1^{(r)}$ as free parameters. The result of this fit is
displayed in the right plot of Fig.~\ref{fig:radii} for $(r_2^{(v)})^2$
and the left plot of Fig.~\ref{fig:mmoment} for $\kappa^{(v)}$ together with
the lattice data and phenomenological results at $\mps = m_\pi$.

The lattice data for the Dirac radius do not seem to agree well with
the ChEFT result. Since it is not clear up to which quark masses the
ChEFT expression is valid, results at even smaller quark masses will be
needed to actually clarify this issue. For both the Pauli radius and
the anomalous magnetic moment the lattice and ChEFT results
look consistent.
It is somewhat surprising that this seems to hold also for rather heavy
quarks. From our preliminary results at very low quark masses
we see first indications for the Pauli radius to bend towards the
phenomenogical value.

Finally, in the right plot of Fig.~\ref{fig:mmoment},
we show our results for the iso-scalar
form factor. From ChEFT a linear quark mass dependence is expected, which is
fully consistent with our lattice calculations.

\section{Conclusions}

We have presented the current status of the calculation of the electromagnetic
form factors of the nucleon by the QCDSF collaboration. At the currently
achieved level of statistical errors, still large uncertainties remain for
the parameterisation of the form factor results.
However, qualitative agreement
with the experimental data has been found, e.g.~the flavour dependence of
the Dirac form factor $F_1$.
As new configurations at very small quark masses are starting to become
available, we are improving our control on the extrapolation of the lattice
results towards the chiral limit. We have found first indications for strong
effects at small quark masses, which have been predicted by ChEFT
calculations.
However, results at even lower quark masses with higher
statistics will be required in order to confirm these predictions.

\section*{Acknowledgements}

The numerical calculations have been performed on the Hitachi SR8000
at LRZ (Munich), the Cray T3E at EPCC (Edinburgh) \cite{Allton:2001sk}
the APE{\it 1000} and apeNEXT at NIC/DESY (Zeuthen), the BlueGene/L at
NIC/FZJ (J\"ulich) and EPCC (Edinburgh). Some of the
configurations at the small pion mass have been generated on the
BlueGene/L at KEK by the Kanazawa group as part of the DIK research
programme.  This work was supported in part by the DFG, by the EU
Integrated Infrastructure Initiative Hadron Physics (I3HP) under
contract number RII3-CT-2004-506078.
We would like to thank A.~Irving for providing updated results for $r_0/a$
prior to publication.

\end{document}